\documentclass[aps,prl,twocolumn,groupedaddress]{revtex4}
\usepackage{amssymb}
\usepackage{amsmath}

\begin{document}

\preprint{MCTP-03-50}
\preprint{DAMTP-2003-128}
\preprint{hepth 0311152}
\title{A Possible Mechanism for Generating a Small Positive
Cosmological Constant}
\author{Gordon L. Kane,$^a$}\email{gkane@umich.edu}
\author{Malcolm J. Perry, $^{a,b}$} \email{malcolm@damtp.cam.ac.uk}
\author{Anna N. \.Zytkow $^{a,c}$}\email{anz@ast.cam.ac.uk}
\affiliation{$^a$Michigan Center for Theoretical Physics, 
Randall Laboratory,  
University of Michigan, 500 E University Ave., Ann Arbor, MI 48109, USA}
\affiliation{$^b$DAMTP, Centre for Mathematical Sciences,
University of Cambridge, Wilberforce Road, Cambridge CB3 0WA, England}
\affiliation{$^c$Institute of Astronomy, University of Cambridge, 
Madingley Road, Cambridge, CB3 0HA, England}
\date{\today}

\begin{abstract}
We argue that in the context of string theory a large number $N$ of
connected degenerate vacua that mix will lead to a ground state with much
lower energy, essentially because of the standard level repulsion of quantum
theory for the wavefunction of the Universe.
We imagine a history where initial quantum fluctuations give an
energy density $\sim m_{susy}^{2}m_{Pl}^{2},$ but the universe
quickly cascades to an energy density $\sim m_{susy}^{2}m_{Pl}^{2}/N.$
\ Then at various phase transitions there are large contributions to the
energy density and rearrangement of levels, followed again by a rapid
cascade to the ground state or near it. \ If this mechanism is correct, the
ground state of the theory describing our world would be a superposition of a
large number of connected  string vacua, with shared superselection sets
of properties such as three families etc.
The observed value of the cosmological
constant is determined in terms of the Planck mass, 
the scale of supersymmetry breaking and
the number of connected string vacua.
\end{abstract}

\pacs{}
\maketitle




Recent observations of type Ia supernovae \cite{a} and
the cosmic microwave background power spectrum \cite{b}
have led to the consensus that there is dark
energy giving a contribution of $\Omega \sim 0.7$ or 
$ \rho_{de} \sim 10^{-10 }\
\mathrm{{eV}^{4}}$ to the total energy density of the Universe.
 This dark
energy can be identified with a positive
cosmological constant, $\Lambda_{de} ,$ 
by $\rho _{de}={\Lambda _{de}/{8\pi G_{N}}%
}.$ $\Lambda _{de}$ is very small on any energy scale in fundamental
physics, being roughly $10^{-122}$ in Planckian units. It is a challenge to
explain this observation\ \cite{c},  particularly without recourse to anthropic
arguments, and using the only quantum theory of spacetime and particle
physics available, namely
string theory. Attempts to understand this problem are hindered by several
events in the history of the Universe 
where a vacuum energy that is like a cosmological constant
is expected to be generated.
As the Universe evolves, there are stages of 
its thermal history where vacuum energy becomes important.
During the inflationary epoch perhaps
$\Lambda \sim m_{unif}^{2},$ with the unification mass scale $m_{unif}$ being 
$\sim 10^{16}\mathrm{GeV}$. Later supersymmetry breaking is
expected to generate $\Lambda \sim m_{susy}^2,$ with 
$m_{susy}\sim 1\mathrm{TeV}.$ At the electroweak 
transition one expects to find $\Lambda \sim
m_{ew}^{2},$ with the electroweak scale being roughly $m_{ew}\sim 100\mathrm{%
GeV}.$ Lastly, there is the QCD confinement transition with $\Lambda \sim
m_{QCD}^{2},$ with $m_{QCD}\sim 250\mathrm{MeV}.$ All of these scales are
very much larger than the scale observed for $\Lambda_{de}$. The cosmological
constant problem(s) require solving at least three logically separate but
probably related issues: (1) The quantum fluctuation contribution to the
energy density $\rho _{o}\sim m_{susy}^{2}m_{Pl}^{2}\sim 10^{80}\mathrm{eV}^4$
must be reduced so
that it is at or below $\rho _{de}.$ (2) The various vacuum energy
contributions need to be similarly dealt with. 
(3) A residual of order $10^{-10}
\mathrm{eV}^{4}$ needs to be explained.

A precursor of our approach is the paper of Yokoyama \cite{d}, who studied
the ground state of a theory with two degenerate vacua mixed by instanton
effects, assuming the cosmological constant vanished in the absolute ground
state. We were stimulated by this treatment of degenerate vacua
to consider what would happen in string theory. Meanwhile,
a number of quite different and  interesting analyses of the 
cosmological constant
problem have appeared in recent years and 
while we have been obtaining the results we 
describe, \cite{e}, \cite{suss}, \cite{feng}, \cite{BP}.

We will formulate our approach in terms of a compactified string theory,
and assume the compact manifold is  Calabi-Yau. This is particular limit of
M-theory rather than being general, and of course we do not yet know whether 
this is  nature's choice. If this approach indeed leads to progress in
understanding the cosmological constant problem(s) there are two 
possibilities. Either the approach can be generalized to other limits of
M-theory, or it is evidence that the compactification is on Calabi-Yau 
manifolds.
Since we compactify on a Calabi-Yau space, the effective 
four-dimensional cosmological constant is expected to be zero before
supersymmetry breaking, \cite{f}.
Our approach  has the advantage of allowing
softly broken $N=1$ supergravity to appear in four-dimensions coupled to
matter. The precise nature of the matter is determined by the properties of
the Calabi-Yau space. Each topologically distinct Calabi-Yau space has a
large moduli space. The absence of massless scalars in nature 
means that quantum
corrections must have arisen that break the moduli space up into
isolated points by the introduction of  some 
potential $V$. $V$ can have many isolated degenerate 
minima at $V_{min}$ corresponding
to many possible physical vacuum states.  However, $V_{min}$ must be
above, or at, the vacuum energy density in the phase
where supersymmetry is unbroken, so that compactification of a Calabi-Yau 
space implies that  all minima are such that  $V_{min} \ge 0$.
Extensive discussion of such a
potential has recently been given by Douglas, \cite{g}.
An expansion of $V$ about
a  minimum is typically taken to be the usual low-energy supersymmetry breaking
superpotential, with
only a finite potential 
barrier between the various
possible vacuum states. A difficulty in making contact with phenomenology
relates to the number of possible vacuum states $N.$ $N$ is thought to be 
enormous
with estimates that range from $\geq 10^{12},$\ \cite{h}  to perhaps more than
$10^{138}$, \cite{g}. The essential thing to keep in mind is that
we probably live in the ground state of the {\it complete} potential,
or at least a very low-lying long-lived state, 
and the properties of
our world are determined by being in this state and not by the
properties of any of the string vacua that might arise in any particular
compactification scheme. This comes about because in systems with large
numbers of minima of $V$, although classically one might live in a single
one of these minima, in quantum mechanics the true eigenstate of
the Hamiltonian has a
wavefunction that is large near many of these minima. This phenomenon is
well established and is responsible for the band structure of solids. Thus
it may well be that the true cosmological constant is small in the ground
state even though it is large if calculated perturbatively about any single
string vacuum, as we argue here.
 
Our argument is basically that mixing of $N$ degenerate 
string vacua by normal quantum mechanical level repulsion leads to 
a ground state with energy density of order the initial one divided by $N$.
The first thing we need to do is to make an estimate of how much the lowest 
level is decreased as a function of the number of connected vacua.
We use a simple one-dimensional quantum mechanical model to illustrate 
this
point. Consider a particle of mass $m$ moving in a periodic potential of
unit period with $V(x)={V_{0}(1-cos(2\pi x))/2}$. This potential has
minima at $x=n$ for any integer $n$. The bottom of each potential well
looks very much like that of a simple harmonic oscillator of angular
frequency $\omega =\sqrt{{2\pi^2 V_{0}}/{m}}$. 
If there was just one such
minimum, then one would expect the ground state energy to be roughly 
${\hbar \omega }/{2}$. However, the true quantum mechanical ground state of
this system does not occur when the particle's wavefunction is
concentrated at the bottom of one of the wells, but rather, by virtue of
Bloch's theorem, one which is non-vanishing in each of the minima of the
potential. The result is that the highly degenerate initial state is
replaced by a band of energies. One can calculate the allowed energy band by
the traditional method of solving the Schr\"{o}dinger equation. However this
problem can also be appoximately solved by using instanton methods.
Consider a set of states $|n\rangle $ that are simple harmonic oscillator
ground states of the particle at the minima $x=n.$ The true eigenstates of the
Hamiltonian are Bloch waves, eigenstates of the translation operator.
A trial Bloch state wavefunction can then be written as 
\begin{equation}
|\theta \rangle =\sqrt{\frac{1}{2\pi }}\sum_{n}e^{in\theta }|n\rangle 
\end{equation}%
where $0<\theta <2\pi ,$ or equivalently theta vacua in the instanton
literature. Since $\langle n^{\prime }|n\rangle =\delta _{n^{\prime },n}$, $%
\langle \theta ^{\prime }|\theta \rangle =\delta (\theta ^{\prime }-\theta ).
$ Consider the matrix element $\mathcal{M}_{\theta ^{\prime
},\theta }=\langle \theta ^{\prime }|e^{-HT/\hbar }|\theta \rangle $, where $%
H$ is the Hamiltonian. As the Euclidean time
parameter $T$ becomes large, $\mathcal{M}%
_{\theta ^{\prime },\theta }\rightarrow \delta (\theta ^{\prime }-\theta
)e^{-E(\theta )T/\hbar }$ where $E(\theta )$ is the energy of the lowest 
Bloch wave
labeled by $\theta $. Were our trial state to contain any other eigenstates
 of the Hamiltonian with higher values of the energy, by taking the 
limit as $T\rightarrow \infty$ their contributions  would be suppressed 
relative to ground state energy,  so 
$E(\theta)$ in this limit is precisely the ground state energy.

The instanton method \cite{i} basically involves rewriting
this expression as $\mathcal{M}_{\theta ^{\prime },\theta }=\sum_{n^{\prime
},n}\langle \theta ^{\prime }|n^{\prime }\rangle \langle n^{\prime
}|e^{-HT/\hbar }|n\rangle \langle n|\theta \rangle .$ In this sum, the
matrix element $\langle n^{\prime }|e^{-HT/\hbar }|n\rangle $ would vanish
unless $n^{\prime }=n$ if the Hamiltonian really were a series of isolated
simple harmonic oscillator potentials. But the real Hamiltonian will allow
the possibility of a particle localized at $x=n$ to tunnel through to $%
x=n\pm 1$. This barrier penetration problem can be solved by the WKB method
and results in $\langle n\pm 1|e^{-HT/\hbar }|n\rangle \sim \langle
n|e^{-HT/\hbar }|n\rangle KTe^{-S_{0}/\hbar }$ where $S_{0}$ is usually
called the instanton action. If one chooses the $+(-)$ sign in this
expression, it is usually referred to as the (anti)-instanton contribution. A
convenient way of thinking of the instanton results from considering the
details of the WKB method. A classical particle can move in the potential $%
-V(x)$ along a path from $x=n$ to $x=n+1$ with zero total energy. This
motion, $\bar{x}=x(t)$ is the instanton. The WKB expression for the
instanton action is therefore $S_{0}=\int_{n}^{n+1}\sqrt{2V(x)}dx$. The
factor $K$ is given by 
\begin{equation}
K=\sqrt{\frac{S_{0}}{2\pi \hbar }}
\sqrt{\bigg|\frac{\mathrm{det}(-\partial_t^{2}+\omega ^{2})}
{\mathrm{det^\prime}(-\partial_t^{2}+V^{\prime \prime }(\bar{x}))}\bigg|}
\end{equation}
In this expression, determinant of an operator means  the
product of all of the eigenvalues of the operator for wavefunctions that
have finite norm and obey appropriate boundary conditions.
The prime in the determinant in the denominator means that the
zero eigenvalue of this operator must be omitted 
because otherwise the expression is ill-defined.
Physically this  corresponds to the location
(in the Euclidean time parameter $T$)
of the center of the instanton. The fact that this center can be anywhere
results in the factor of $T$ in the matrix element.
The factor of $\sqrt{\frac{S_0}{2\pi\hbar}}$ comes about
because one has to omit the zero mode from the determinant in the denominator
of $K$.
One should notice that $K$ has the dimension of an inverse length,
coming from the fact that there is one fewer eigenvalue in the determinant
in the denominator relative to the numerator.
$K$ represents the difference between quantum fluctuations in the harmonic
oscillator ground state and those associated with the instanton tunneling
event. 

The above expression describes what happens if the particle tunnels once from
$ |n\rangle $ to $|n+1\rangle $, but it is clear that one must allow the
possibility that the particle can tunnel an arbitrary number of times. This
is just done by supposing that one adds the contribution of $p$ instantons
and $q$ anti-instantons with $p=q+1$ to indicate
that one has only jumped by one step. To calculate a more general
matrix element where one jumps from $n$ to $n^{\prime }$ one adds the
contributions from $p$ instantons and $q$ anti-instantons subject to 
$p-q=n^{\prime }-n.$ This results in the expression 
\begin{align}
\mathcal{M}_{\theta ^{\prime },\theta } &=\sum_{n,n^{\prime },p,q}\langle
\theta ^{\prime }|n^{\prime }\rangle \frac{1}{p!}\frac{1}{q!}%
(KTe^{-S_{0}/\hbar })^{p+q}\delta _{p-q,n^{\prime }-n}\nonumber\\
&\times \langle n|\theta \rangle e^{-{\frac{\omega T}{2}}}  \nonumber \\
&=\frac{1}{2\pi }\sum_{n,n^{\prime },p,q}\frac{1}{p!}\frac{1}{q!}%
(KTe^{-S_{0}/\hbar })^{p+q}e^{in\theta -in^{\prime }\theta ^{\prime }}e^{-{%
\frac{\omega T}{2}}}  \nonumber \\
&=\sum_{p,q}\frac{1}{p!}\frac{1}{q!}(KTe^{-S_{0}/\hbar
})^{p+q}e^{-i(p-q)\theta^{\prime }}\delta (\theta -\theta ^{\prime })e^{-{%
\frac{\omega T}{2}}}  \nonumber \\
&=e^{2KT\cos \theta ^{\prime }e^{-S_{0}/\hbar }}\delta (\theta -\theta
^{\prime })e^{-{\frac{\omega T}{2}}}
\end{align}%
where we have used $\delta (\theta -\theta ^{\prime })={1/{2\pi} }%
\sum_{n}e^{-in(\theta -\theta ^{\prime })}.$ The factors of $p!$ and $q!$ in
the sum over the (anti)-instantons arise because the (anti)-instantons are
indistinguishable. By comparing this expression with the definition of $%
E(\theta )$ we see that 
\begin{equation}
E(\theta )=\frac{\hbar \omega }{2}-2\hbar K\cos \theta e^{-S_{0}/\hbar }
\end{equation}%
This shows that the lowest energy state, corresponding to $\theta =0,$ is
depressed relative to the simple estimate by an amount $2\hbar
Ke^{-S_{0}/\hbar }.$ It should be noted that the band is a continuum of
width $\Delta E=4\hbar Ke^{-S_{0}/\hbar }$ whose wavefunctions correspond to
varying $\theta $ from $0$ to $\pi $. The fact that this is a continuous
band is a consequence of allowing an infinite number of minima in the
potential. Were one dealing with the case in which there were a finite
number, $N$, of such minima, as we expect in string theory, then the
locations of the boundary of the band would be unaltered for $N\gg 1$ but
instead of a continuum there would be $N$ discrete energy levels in the band
each separated by $\sim {{\Delta E}/{N}}.$ 

The nature of the instanton calculation is such that it can be extended in a
natural way to the case of field theory. The basic formalism is unaltered
except that the energy of a state is now replaced by the energy density, and
in the expression for $K$ the determinant factors are replaced by the
corresponding determinants of the wave operators governing the fluctuations
in the quantum fields. The replacement of energy by energy density in field 
theory comes about because the basic degrees of freedom in field theory are 
oscillators at each point in space rather than single particles as in
quantum mechanics.

The semi-classical effective 
action for gravitation coupled to a collection of scalar fields $\phi_i$
is
\begin{equation}
S=\frac{1}{16\pi G_N}\int g^{1/2}d^4x (R-2\Lambda) + \int g^{1/2}d^4x
\bigl(\sum_i \partial \phi_i^2 +V(\phi)\bigr)
\end{equation}
where $R$ is the Ricci scalar, and $V(\phi)$ is the potential for the scalars.
Note that the only coupling constant in this expression is the 
$\frac{1}{16\pi G_N}$ in front of the gravitational piece of the action.

The situation we are discussing in cosmology is analagous to
the quantum
mechanical setup we just described, except for the physical interpretation
of the instantons.
We are assuming that in the absence
of supersymmetry breaking,
no vacuum energy is generated. Once supersymmetry is broken, 
there is a potential that governs which
of the Calabi-Yau spaces we are in. \ It has many minima. \ Tunneling
removes the degeneracies, pushing one level up and the other down
repeatedly.  Quantum fluctuations  give rise naturally to a vacuum
energy of $m_{susy}^{2}m_{Pl}^{2}$ which is the field theory analog of
the zero point energy of the harmonic oscillator in our quantum mechanical
example.
This result will be modified by instanton effects because of the
possibility of Hawking-Moss 
tunneling to any of the other possible vacua. Consider the
case where all quantum corrections are turned off so that one is dealing
with classical field theory. The possible vacua are then described by the
product of four-dimensional Minkowski space with any Calabi-Yau space. No
restriction is put on either the differential structure or the topology of the
Calabi-Yau space. One knows that the differential structure is
completely free, but it transpires that in string theory, one can even
change the topology of the Calabi-Yau  space without causing any
singularities that would render such a change inadmissible, \cite{pc}.
When quantum
corrections are turned on, some potential develops leading to $N$ possible
discrete vacua. Just as in our one-dimensional quantum mechanical example,
eigenstates of the Hamiltonian will not be concentrated only in a single
minimum since there are a huge number of minima all with the same vacuum
energy at the bottom of their potential wells. It is perhaps rather
premature to make a detailed model of what the actual potential looks like,
but we can make some general observations. We
expect that our results will applicable whatever the actual nature of the
potential is for more or less the same reasons that the band structure of
solids is more or less independent of the material.

In the overall picture we have a large number of minima, 
\lq\lq vacuum states.\rq\rq\  There will be a complicated
energy surface between these minima over which one could travel if one
had sufficient energy available. Classically, if one does not have 
the energy available,
one is trapped in a particular minimum and all of the other minima are 
irrelevant. Quantum mechanically however the 
effect of other discrete minima is extremely important. One can tunnel between
various of these minima and the rate at which this happens has a profound
influence on the physical consequences of this possible mixing of the vacua.
To say something definite about the tunneling rates requires some
understanding of  this potential energy surface. At the bottom
of each potential well, one expects that the potential is 
given by the usual low-energy supersymmetry breaking  superpotential. 
Thus we expect that the curvature
of the potential energy function at the bottom of each well is governed 
essentially by the supersymmetry-breaking scale, $m_{susy}.$ More precisely,
the relevant scale should be proportional to $m_{susy},$ so that 
it vanishes with $m_{susy},$ but it could depend on other physics too.
The situation we
are describing is a little bit like looking at the surface of a lake.
Prior to supersymmetry breaking, the surface of the lake is completely
flat. Once supersymmetry breaking is turned on,  the surface of
the lake becomes irregular with all
of the surface features  governed by a single scale.
Under these circumstances, the only scale relevant to instanton processes
will be $m_{susy}$. If we were thinking in the more traditional way where
the observed universe corresponds to being confined to a single 
minimum, this could be a catastrophe since the potential
barriers between different minima would be relatively small, and so
the universe could quickly tunnel to some new minimum where the
physics was quite different to the original minimum. Such a  transition
from one state to another is rather like that encountered in \lq\lq old\rq\rq
\ inflation, and could lead to regions of new vacuum expanding at 
the speed of light into a region of the old vacuum. Such an event 
has not happened 
even on cosmological timescales. However, in our picture, this is not important
since the universe we are considering is a superposition of states 
described by all accessible vacuum states. No possibility of
a sudden tunneling to a world unlike the one we now occupy exists.

 It is 
possible that the potential also involves $m_{Pl}$ in some essential way,
though this seems rather unnatural,
or perhaps the curvature is determined by the Hubble parameter $H,$
since supersymmetry is broken by the energy density in the early universe 
\cite{DRT}.
Thus 
we will momentarily allow for these more general cases. 
Now we apply Hawking-Moss instanton methods to evaluate the spectrum and 
energy 
density of the ground
state energy band. For definiteness, suppose that the actual potential
consists of a $d$-dimensional hypercubic lattice of minima at zero vacuum
energy at the points in field space $(n_{1},\dots ,n_{d}).$ 
The minima are separated by
barriers of height roughly on the scale $M$. As we outlined above, it
is not initially clear
what scale $M$ should have.
The widths of these potential barriers are presumably
controlled by the same basic scale. However, in the absence of supersymmetry
breaking, the potential is flat, and supersymmetry breaking is what gives 
rise to the curvature of the potential. So, we expect that 
$H \gtrsim M \gtrsim m_{susy}.$
Equally unclear is the value of $d.$
It is determined by which minima can communicate with which other minima via
instantons, so all we can be sure of is that $1< d< N.$ 

For
each direction, we can introduce a separate $\theta $ variable, so that our
states are described by $(\theta _{1},\ldots \theta _{d}).$ Let 
\begin{equation}
\mathcal{M}_{\theta _{1}^{\prime },\ldots \theta _{d}^{\prime },\ \theta
_{1}\ldots\theta _{d}}=\langle \{\theta _{i}^{\prime }\}|e^{-HT/\hbar }|\{\theta
_{i}\}\rangle 
\end{equation}%
where we now evaluate this expression in a box of volume $V$. As $T$ becomes
large, then $\mathcal{M}\rightarrow \prod_{i}\delta (\theta _{i}^{\prime
}-\theta _{i})e^{-\rho (\{\theta _{i}\})VT/\hbar }$ where $\rho (\{\theta
_{i}\})$ is the energy density of the lowest energy states described by $%
\{\theta _{i}\}$. Then by the usual type of instanton argument, we find 
\begin{align}
{}&\mathcal{M}_{\theta _{1}^{\prime },\ldots \theta _{d}^{\prime },\ \theta
_{1}\ldots\theta _{d}}=\sum_{\{n_{i}^{\prime
},n_{i},p_{1},q_{i}\}}\prod_{\{p_{i},q_{i}\}}\frac{1}{p_{i}!}\frac{1}{q_{i}!}
\nonumber\\
&\times (\mathcal{K}TVe^{-S_0/\hbar})^{p_i+q_i}
\prod_{\{n^\prime_i,n_{i}\}}e^{in_{i}
\theta _{i}-in_{i}^{\prime }\theta _{i}^{\prime
}}e^{-\rho _{o}VT/\hbar }
\end{align}%
Here we have used the appropriate field theory extrapolation of the
instanton formula  we used in the quantum mechanical model.
The principal difference is that one is now calculating the energy
density of the vacuum and the determinant factors in  $\mathcal{K}$ are now 
replaced by the ratio of
functional determinants describing the quantum fluctuations of the fields in
the model.
Just as in the quantum mechanical case, we
evaluate $\mathcal{M}$ giving 
\begin{eqnarray}
\mathcal{M}_{\theta _{1}^{\prime },\ldots \theta _{d}^{\prime },\ \theta
_{1}\ldots\theta _{d}}&=(\prod_{i}\delta (\theta _{i}^{\prime }-\theta
_{i}))\nonumber\\
&e^{2\sum_{i=1}^{d}\mathcal{K}VT\cos \theta _{i}e^{-S_{0}/\hbar }}e^{-\rho
_{o}VT/\hbar }
\end{eqnarray}%
where $S_{0}$ is the action of the instanton. Thus, a general expression
for the vacuum energy density $\rho$ 
for a state described by $\{\theta_i\}$ is
\begin{equation}
\rho(\{\theta_i\}) = \rho_0 - 2\sum_i^d \mathcal{K}
\cos\theta_i\ e^{-S_{0}/\hbar}.
\end{equation}
Thus the lowest energy density
in the band $\rho _{min}$ is found by choosing all the $\theta _{i}$ to
vanish given by 
\begin{equation}
\rho _{min}=\rho _{o}-2\mathcal{K}de^{-S_{0}/\hbar }
\end{equation}%
This expression makes it look as if $\rho_{min}$ could be negative. On
general grounds, we know that $\rho_{min}\geq 0,$ so  $\rho_{min}<0$
would mean  that our approximations have broken down. Equation (10) is our
main technical result.

We now need to describe the precise nature of these instantons and
what they do and do not  connect.
Each instanton describes the amplitude of tunneling between one
particular minimum of the potential and one that is adjacent to it 
in field space, that is between, for example $(n_1,\ldots,n_i,\ldots,n_d)$
and    $(n_1,\ldots,n_i+1,\ldots,n_d)$. The situation we are describing here
is quite different from the quantum mechanical example. The minima of the 
potential do not correspond to points in physical space, but rather describe 
the particular Calabi-Yau space on which ten-dimensional spacetime is 
compactified. In the traditional way of thinking about compactification, this 
corresponds to a particular vacuum state in which the wavefunction of the 
universe is concentrated. The precise nature of the Calabi-Yau space would 
then determine the physics of the low-energy world (particle spectrum, 
coupling constants, masses etc.,) {\it if} the wavefunction could be so 
localized. But, the wavefunction must spread out over all of these vacua
or at least some superselection sector. We expect that the wavefunction of
the universe will therefore be some superposition of many of these 
Calabi-Yau vacua, and then presumably low-energy physics is determined
statistically from the properties of these many vacua.
A potential question then is why must one consider such a 
superposition, analogous to the theta vacua of QCD? If there were a conserved
current, then states would be labeled by the corresponding charge. 
However, suppose that one imagined 
starting  off
 the universe with the wavefunction concentrated in one particular minimum,
and the Hamiltonian were such that the wavefunction could leak into
another distinct minimum. Then having the wavefunction concentrated
in one specified minimum could not possibly be an eigenstate of the 
Hamiltonian. To find eigenstates of the Hamiltonian one must include the 
possibility that one has a superposition of various states making
up a wavefunction with support in potentially a considerable number of distinct
minima. 

{\it Thus the transitions are not realized 
in spacetime as
tunneling from one metastable state to another, but rather 
as a mechanism for calculating the energy spectrum of the theory taking into 
account these effects which can never be calculated in perturbation theory.}
This is to be contrasted with the usual uses of instantons in which 
they provide a mechanism by which the universe can start in some particular 
vacuum state and evolve, by a quantum mechanical tunneling process, to a new 
vacuum. Examples of this process include \lq\lq old \rq\rq inflation,
\cite{Guth} and
studies of the cosmological constant problem in most of the papers we have 
been able to find. In particular, the most common procedure whereby
a bubble of \lq\lq true \rq\rq vacuum nucleates in a sea of  \lq\lq false 
\rq\rq vacuum would not be expected to give the correct answer for the 
problem we are considering. 

There are two types of instanton that have been used to evaluate 
such amplitudes. The first (historically) is the Coleman-De Luccia
\cite{CDeL} instanton
which is not appropriate for us. In the absence of gravitation, their 
instanton has a size that is inversely
proportional to the difference in the classical energy densities
between the two adjacent minima.
Since  the minima are degenerate for purposes of calculating the
spectrum, the difference is zero and so the size
must be very large. So we believe that this instanton would have a physical
dimension that is potentially greater than the size of the universe. 
The universe we are describing is de Sitter, since the effective cosmological 
constant is, and will remain, positive according to our results. 
It has a spatial extent that is 
finite. So their instanton is not  relevant. Although this picture is 
modified by the inclusion of gravitation, the modifications are slight and
do not affect the above conclusion \cite{CDeL}, \cite{HM}.

The second type is the so-called Hawking-Moss \cite{HM} instanton. 
Here, the values
of the fields are constant in space and  such that one sits at the top of 
the potential barrier  between adjacent vacua. Thus one expects the 
energy density there to be $M^4$.
If one were in infinite 
flat space,
the resultant action $S_0$ for this instanton would be also be infinite, and
the corresponding amplitude would vanish. But, since we are in de Sitter space,
with finite spatial volume, the contribution is finite, and the amplitudes
are  not suppressed. In fact, we can  estimate the action for 
this instanton. The potential term in the action we denote by $V(\phi)$
where $\phi$ are the relevant collection of fields, such as moduli fields 
or any fields that parameterize the fluctuations. The Euclidean action
for constant $\phi$ is
\begin{equation}
S=\hbar \int g^{1/2}d^4x V(\phi),
\end{equation}
where the integral is taken over four-dimensional Euclidean de Sitter space.
Since $V(\phi)\sim M^4$ and the volume of Euclidean de Sitter space is 
$H^{-4}$,
the action $S_0 \sim \hbar M^4 H^{-4}$, which is $\lesssim 1$ 
at the supersymmetry
breaking scale or earlier.
These instantons occur, and are relevant,
whenever there is a finite potential barrier between any minima in a
spacetime of finite volume..
These instantons are the ones that are relevant to the calculation of the 
spectrum of energies of the superposition of degenerate string theory 
vacua we are discussing.

Next, we need to determine $\mathcal{K}$. $\mathcal{K}VT$ can be evaluated by 
standard Feynman diagram type calculations. To one loop, it is given by the
ratio of the determinants of those operators governing the fluctuations
of the quantum fields in our problem evaluated in the presence of 
the instanton field to those evaluated without. There is a large literature
on such objects, see for example \cite{GP}, \cite{CD}, \cite{FT}.
The determinants in the absence of the instanton depend only on
$H$ and the curvature at the bottom of the potential well, whereas
in the presence of the instanton they depend on $H$ and the curvature 
at the top of the barrier. Thus we expect that roughly $\mathcal{K}\sim H^4$
in the early universe.
 
Since the levels are generated by splitting of degenerate
levels mixed by instanton effects, the lowest level can never become
negative. \ It must always lie above the positive (or zero)
minimum of the potential.  If
each vacuum state could communicate with only a single nearest neighbor,
then $d\sim 1$. \ If all vacua can rapidly tunnel to the others, then 
$d\sim N.$ We expect 
$d\lesssim N,$ where $N$ is an effective number of string vacua that are
connected. The separation between the
various states in this band is $\sim {{\mathcal{K}d}/{N}}e^{-S_{0}/\hbar }.$

Suppose that $\mathcal{K}de^{-S_{0}/\hbar }$ is of order $\rho _{o}.$
Then the bottom
of the band will reach down to a vacuum energy of nearly zero. Since the
separation of the energy densities is roughly
 $(\mathcal{K}d/N)e^{-S_{0}/\hbar }\sim
\rho _{o}/N,$ we can expect that $\rho _{de}\sim \rho _{o}/N.$ For this to
work as observed, since $\rho_o \sim m_{susy}^2m_{Pl}^2$, $S_0/\hbar \sim 1$
and 
$\mathcal{K}\sim H^4\gtrsim m_{susy}^4$,  we find
$d\lesssim \frac{m_{Pl}^{2}}{m_{susy}^2}
\sim 10^{32}e^{S_{0}/\hbar }$ and 
$N\sim \frac{\rho _{o}}{\rho _{de}}\sim 10^{90}$. 
For $M \sim m_{susy},$ as
discussed above, we find 
$d\sim 10^{32}$.  
Since $d\ll N,$ we expect our approximation to be reliable. 
(If $H$ is relevant rather than $m_{susy}$ then $d$ might be somewhat smaller.)
Rather amazingly, these figures do not appear to contradict any known
facts or estimates about string theory compactification. 
String theorists should regard
this as an experimental determination of the value of $N$. Cosmologists
should perhaps regard this as a prediction of the value of $\Lambda_{de}.$

In this picture, the universe is not in a state described by any one minimum
of the effective potential. This is in accordance with what one expects from 
generic quantum mechanical systems with large numbers of degenerate
minima. Whilst this is not alarming for the cases of something like an
electron in a crystal, it is a bit strange within the context of
cosmology. The wavefunction of the universe would seem to be described by
the statistical properties of many minima. However $d \ll N$ so it seems
as if the actual number of states that are involved is relatively small,
only $10^{32}$ or so.
One would expect  that these states would all have features in common with 
the observed universe, such as there being three families, a low-energy
gauge group of $SU(3)\otimes SU(2) \otimes U(1),$ massless quarks and leptons,
etc. and would have the same values of all conserved quantities. Such 
quantities would be a sort of superselection set.

Finally, as discussed at the beginning of the paper, it is necessary to deal
with not only the energy from the quantum fluctuations but also the repeated
release of energy density as the universe cools. \ What we expect
qualitatively is that the relaxation time for the universe to cascade to the
lowest levels is small, because the levels are closely spaced and 
this old-fashioned
tunneling is rapid. \ Initially the energy from quantum fluctuations gives $%
N$ levels with energy density $\sim m_{susy}^{2}m_{Pl}^{2}.$ \ Then
the mixing splits the levels and the universe falls down to the low-lying
levels with energy
density $\sim m_{susy}^{2}m_{Pl}^{2}/N.$ \ Inflation and reheating
are occurring and the universe cools. \ At various phase transitions
additional energy density is dumped in, rearranging the levels. \ Quickly
the tunneling splits them, again leaving the universe in a low level.
For each case $N, d$ will be different: $N^\prime, d^\prime$ etc.

Perhaps protein folding is a useful analogy. 
Think of the universe as a protein. Initially the protein is in a
high energy level, and then it quickly falls mainly by tunneling 
to a minimum of
the energy. \ The physics of protein folding has for some time been
described in terms of an \lq\lq energy landscape\rq\rq, a terminology recently
introduced into \ particle physics in \cite{suss}.
For a review of the situation in
protein folding see \cite{j}.

For this picture to be valid it is necessary first that the
cascade take place sufficiently rapidly. 
This is much easier to achieve if the $N^\prime$
levels have widths at least of the order of their splitting.
Although at present we are unable to estimate the width of these
levels, because there are so many levels this seems entirely plausible.
There must also  
be a mechanism for dissipating the released energy. \ Since
vacuum energy couples to gravity, surely graviton emission will be an
important way to carry the energy away. That is because $only$ the
gravitational field couples to the vacuum energy density.  So we expect that
graviton emission will cause $\theta_i$ to decay. Because of gauge invariance,
it turns out that graviton pair production is the most important process, 
so that gravitons are produced with opposite momenta and carry away energy
in the usual way, \cite{Ford}.
Perhaps it is possible for other processes also to contribute,
for example the production of branes as has been considered in 
\cite{k}, \cite{feng} 

To estimate such effects, we suppose that
the $\theta _{i}$ can be viewed as fields with spatial and time dependence.
If that was all there was to it, then the equations of motion for $\theta
_{i}$ would imply that $\theta _{i}$ were constant. That is because
vacuum energy
couples to the gravitational field through a term in the 
four-dimensional tree level action of the form 
\begin{equation}
(\det {g_{ab}})^{1/2}\rho(\{\theta_i\}).
\end{equation}%
Quantum gravity effects will however  generate  additional 
terms in the one-loop effective 
Lagrangian,  $\sim m_{Pl}^{2}\sum_{i}\partial \theta _{i}\partial
\theta _{i}$ and $\sim m_{Pl}^4\sum_{i} \theta_{i}^2$.
The first  is the standard kinetic energy expression
for the fields $\theta_i$, and the second is just the usual mass term.
The resultant semi-classical equations of motion then can be 
derived from the effective Lagrangian for each $\theta_i$,
which leads to the
decay of $\theta_{i}$'s to zero on a Planck timescale,
rapidly enough to guarantee that the $\theta_i$
are minimized along with the energy density.

If correct, our approach has a \ number of phenomenological implications. \
Since the level splitting mechanism can push the lowest levels down but not
below the absolute minimum of the potential, a small non-zero cosmological
constant is a generic result. \ Since the energy density is a cosmological
constant, we predict a time-independent equation of state, $w=-1.$

The \lq\lq why now \rq\rq  problem
 --- why is the size of $\Omega _{de}$ about the same
as $\Omega _{dm}$ so the dark energy is dominating now rather than much
earlier, in which case it could have prevented formation of gravitationally
bound systems --- is in principle solved by our approach, if
the number of connected string vacua and the fluctuation energy density could 
be calculated. Of course, in practice that is very difficult.
If the number of connected string degenerate minima is very large, then
the number of possible ground states with energies of order $10^{-3}$ eV or
less is large. The actual ground state of a universe can be any of the low
lying ones, many of which would allow solar system and galaxy formation.
This requires $N\gtrsim m_{susy}^{2}m_{Pl}^{2}/10^{-10} \mathrm{eV}^{4},$
which is consistent with our  estimates. But to fully explain 
\lq\lq why now \rq\rq it is necessary actually to calculate 
$\rho_0$ and $N$ rather accurately.

 While our approach is 
not an anthropic one, a residual anthropic element could enter
until it could be proved that the number of connected string vacua was
large enough to guarantee that there were many very low lying levels, 
and that the relaxation mechanism was very efficient in tunneling to
very low lying levels. All universes that can form structure will --- there
is no reason why we should live in a highly probably one.


A persistent problem for supergravity and string-based model building is
what to do about the cosmological constant. \ Our approach suggests that the
solution to the cosmological constant problem will not impact significantly
other aspects of models, such as soft breaking Lagrangian parameters or
Yukawa couplings in the superpotential or CP violation properties. Our
approach may also offer the possibility of changes in the low-energy effective
couplings at phase transitions. For example near the 
electroweak or QCD transitions changes could occur since then 
additional vacuum energy poured into the universe from the phase transitions
could trigger vacuum readjustments that could allow for changes in scalar 
vacuum expectation values at these eras.

An
apparently unpleasant consequence of our approach is that our world is not
in the minimum of the potential of a single compactification but rather a
superposition  of a large number of different ones. \ Of course the relevant number
that mix must all have the same value of the number of families and any
similar properties, so the total number of string vacua can be considerably
larger than the connected ones.  Probably all of the properties of the
world that are shared by all of the connected states, such as the number of
 families, the standard model gauge group, softly broken $N=1$ supersymmetry, 
etc.,
can be derived even though the ground state is a superposition.
More generally, it will not only be hard to identify the correct string
vacuum from first principles, it is probably both not possible and not
relevant. \ In the end that may be a virtue --- just as inflation can
provide powerful explanatory power (once the inflaton is identified), so
too perhaps having the universe in a highly mixed state will allow progress
once it is better understood.

One might start to wonder when this mechanism for producing 
 a cosmological constant becomes the dominant one. At very early times
after compactification, there must surely be a period of inflation
at or around the GUT scale. This means that whatever corrections to the 
cosmological constant we generate must be small compared to this rather larger
 scale. This puts a somewhat different limit on the value of $N$.
By analogy with our previous estimates, we find that $N << 
\frac{\rho_{unif}}{\rho_{de}}\sim 10^{116}$.
At later stages, periods of inflation
could take place, but in the end we will always relax down
to close to the bottom of the band of allowed energies. Thus, for example it 
could be that there is inflation at the electroweak transition. Equally, it
might happen that the precise state we end up in before and after this
period of inflation could be different, leading to (hopefully not drastic)
differences in low-energy physics before and after the transition. 
Similar comments also of course apply to the QCD phase transition. 
In contrast, at the supersymmetry breaking transition, we first 
find our mechanism to give interesting results leading to a much lower
cosmological constant than naively envisaged.

What we have presented is a mechanism that seems to be able to deal with
the cosmological constant problem(s). To establish that it indeed does, it will
be necessary to have a much better understanding of the number of connected
string vacua, the relaxation issues and the interactions connecting
different vacua and perhaps other issues. We are presently examining 
these questions further, \cite{KPZZ}.
 If this 
approach is indeed correct, then the dark energy should just be thought of 
as an energy density purely due to ubiquitous quantum fluctuations. 
\bigskip 

We would like to thank Michigan Center for Theoretical Physics for its 
generous hospitality to MJP and ANZ, and the US Department of Energy for
partial financial support,  S. Krimm for providing helpful references
about protein folding, and D. Berman, M. Douglas, G. Gibbons, F. Larsen, 
J.T. Liu, R. McNees, J. Wells and R-J. Zhang for helpful conversations.

\end{document}